\documentclass[%
 reprint,
preprintnumbers,
nofootinbib,
amsmath,amssymb,aps,superscriptaddress
]{revtex4-2}
\counterwithout{equation}{section}
\usepackage{amsmath,float,graphicx,placeins,amssymb,multirow,pgfplots,pgfplotstable}
\usepackage{empheq}
\usepackage{booktabs}
\usepackage{tikz}
\usepackage{amsmath}
\usepackage[most]{tcolorbox} 
\usepackage{amsmath,amssymb,amsfonts,bm}
\usepackage{mathtools}
\usepackage{physics}
\usepackage{microtype}
\usepackage{xcolor}
\usepackage{comment}
\usepackage{enumerate,slashed,cancel,comment,color,bbm,graphicx,rotating,placeins,mathtools,multirow,hhline}
\usepackage[normalem]{ulem}
\usepackage{array}
\usepackage{hyperref}
\usepackage{color}
\usepackage{booktabs}
 \hypersetup
 {
   colorlinks,
   linkcolor={blue!80!black},
   citecolor={blue!70!black},
   urlcolor={blue!70!black}
 }

\usepackage{hyperref}
\usepackage{tabularx}

\newcommand{\be}{\begin{equation}}
\newcommand{\ee}{\end{equation}}
\newcommand{\bea}{\begin{eqnarray}}
\newcommand{\eea}{\end{eqnarray}}
\newcommand{\ba}{\begin{aligned}}
\newcommand{\ea}{\end{aligned}}

\newcommand{\nn}{\nonumber\\}

\pgfplotsset{compat=1.18} 
\begin{document}
\preprint{KCL-PH-TH/2025-46}
\preprint{TUM-HEP-1578/25}
\title{Exact Renormalisation Group Evolution of the Inflation Dynamics:\\
Reconciling $\alpha$-Attractors with ACT
}

\begin{abstract} 
We present a non-perturbative framework for the dynamics of slow-roll inflation that consistently incorporates quantum corrections, based on an alternative functional renormalisation group (RG) approach. We derive the coupled Friedmann–RG flow equations governing the joint evolution of spacetime, the inflaton field, and its effective potential. Applying this formalism to $\alpha$-attractor E-models, we find that the RG flow induces a dynamical destabilisation of the inflationary trajectory, leading to a premature termination of slow roll. Remarkably, the resulting predictions bring $\alpha$-attractors into full agreement with the latest ACT data without introducing new physics beyond a consistent quantum-corrected treatment of the inflaton dynamics.
\end{abstract}

\author{Jean Alexandre}
\email[Email address: ]{jean.alexandre@kcl.ac.uk}

\author{Lucien Heurtier}
\email[Email address: ]{lucien.heurtier@kcl.ac.uk}

\affiliation{Theoretical Particle Physics and Cosmology, King’s College London,\\ Strand, London WC2R 2LS, United Kingdom}

\author{Silvia Pla}
\email[Email address: ]{silvia.pla-garcia@tum.de}

\affiliation{Physik-Department, Technische Universit\"at M\"unchen, James-Franck-Str., 85748 Garching, Germany}

\maketitle

\paragraph*{{\bf Introduction ---}}
Since it was suggested in the 80s~\cite{Guth:1980zm,Starobinsky:1980te,Linde:1981mu}, the cosmic inflation paradigm, in which a scalar field dubbed the \mbox{{\em inflaton}} rolled down a relatively flat potential in the early universe, has passed successfully many of the observational tests (see e.g.~\cite{Baumann:2022mni} for a review). In particular, single-field slow-roll inflation models  explain remarkably well why the primordial spectrum of scalar perturbation observed in the cosmic microwave background (CMB) by {\it WMAP}~\cite{WMAP:2010qai} and {\it Planck}~\cite{Planck:2018vyg} is nearly flat. Moreover, searches for tensor modes in the CMB from BICEP/Keck~\cite{BICEP:2021xfz} tightly constrained inflationary scenarios, ruling out vanilla models of chaotic inflation, and favouring concave {\em plateau}-like inflation potentials such as in Starobinsky~\cite{Starobinsky:1980te}, or more generally $\alpha$-attractor~\cite{Kallosh:2013yoa,Kallosh:2014rga,Kallosh:2013maa} scenarios. Recently, high-resolution cosmic microwave background (CMB) observations from the Atacama Cosmology Telescope
(ACT)~\cite{ACT:2025fju} and the South Pole Telescope (SPT-3G)~\cite{SPT-3G:2025bzu}, provided the community with a new window to constrain primordial cosmology~\cite{Fairbairn:2025fko,Ellis:2025AttractorReheating,Drees:2025RefinedStarobinsky,Berera:2025ACTWarm,Pallis:2025PalatiniACT,Wolf:2025AttractorsRadiative,Addazi:2025CurvatureCorrectionsACT,LiuYiGong:2025HiggsReheatingACT,Maity:2025NonBD_ACT,GaoGongYiZhang:2025NonMinimalACT,Yuennan:2025RadiativeHiggsACT,McDonald:2025HiggsVLQ_ACT,BroadberryHookMondal:2025WarmPseudoScalar,AokiOtsukaYanagita:2025HeavyFieldsACTSPT,PengPiao:2025Oscillations_ACT_SPT3G,HaquePalPaul:2025ACTDR6AttractorsReheating,HaquePalPaul:2025HiggsStarobinskyACT}. In particular, data from ACT~\cite{ACT:2025tim} favours a spectral index $n_s =  0.974 \pm 0.003$, which is on the edge of excluding $\alpha$-attractor scenarios at 95\% of confidence level. 

Reconciling $\alpha$-attractor scenarios with ACT so far required invoking additional new physics leading to a distortion of the $\alpha$-attractor potential, effectively pushing the spectral index to slightly higher values around the pivot scale~\cite{Ellis:2025AttractorConstraints,Haque:2025ACTAttractorsReheating,Yi:2025ACTReconstructionPolynomialAlpha,Heidarian:2025GUPAlphaAttractorACT,Chakraborty:2025WarmAlphaAttractors,Frolovsky:2025OneLoopETypeAlpha}. One interesting such possibility is to consider the running of the inflation potential induced by the coupling of the inflaton field to light fermions~\cite{Ellis:2025bzi} with a coupling tuned to accommodate the ACT data. In this context, like in numerous works incorporating the effect of quantum corrections on the inflationary trajectory~\cite{Albrecht:1982zz, Dvali:1994ms, Stewart:1996ey, Covi:1998jp, Covi:2004tp, Shafi:2006cs, Okada:2013BminusL, Barenboim:2013wra, Heurtier:2019eou}, renormalisation is treated perturbatively, and in a quasi-static manner in which the renormalisation scale of the theory evolves very slowly with the slow-rolling field, deep inside the quasi-de Sitter era.

In this {\em letter}, instead, we consider the inflation self-interactions only and explore how quantum corrections may affect the inflationary dynamics beyond the leading order slow-roll approximation. To do so, we use non-perturbative {\em functional renormalisation} techniques based on the Wetterich average effective action ~\cite{Wetterich:1989xg,Wetterich:1992yh,Wetterich:2001kra}, which allows us to track the evolution of the inflaton potential while the Hubble scale decreases and quantum perturbations exit the horizon and do not contribute to the renormalisation flow. In the context of non-perturbative functional renormalisation, several works have implemented spatial coarse-graining on de Sitter or FLRW backgrounds, finding scale-dependent effective potentials, interesting nonperturbative infrared phenomena, or explicit dependence on the underlying quantum state \cite{Serreau:2013eoa,Guilleux:2015pma,Guilleux:2016oqv,Kaya:2013bga,Banerjee:2022xvi} (see also Ref. \cite{Moreau:2018ena} for backreaction effects on fixed de Sitter). Complementarily, open-system, in-in and stochastic approaches have captured how dissipative and noise effects emerge in the long-wavelength inflationary dynamics by integrating out short-wavelength fluctuations \cite{Calzetta:1999zr,Boyanovsky:2005sh,Prokopec:2017vxx}. 

Here, we take a qualitatively different approach: we derive a new master equation that couples the Hubble-scale–dependent flow of the inflaton potential to the Friedmann dynamics that dictates the dynamics of the classical background field.  As we will see, this evolution can drastically affect the inflationary dynamics towards the end of inflation, considerably changing the field value at which inflation ends, and thus the location of the potential where the inflation observables at horizon exit need to be computed.

\vspace{5pt}
\paragraph*{{\bf Time-Dependent Coarse-Graining in Space ---}}
We consider the Friedmann-Lemaitre-Robertson-Walker (FLRW) metric $ds^2=g_{00}dt^2+a^2(t)(d\vec x)^2$, 
and for each time slice, we perform a coarse graining in space by integrating out ultraviolet (UV) Fourier modes of the inflaton. 
The physical energy scale providing a natural separation between UV and infrared (IR) modes is the Hubble rate $H$, whose covariant expression 
with respect to time redefinition is
\be\label{defH}
H=\frac{1}{\sqrt{-g_{00}}}~\frac{\dot a}{a}~,
\ee
which is proportional to the extrinsic curvature for FLRW spacetime. The resulting effective description is provided by the potential $U_H$,
for which we derive a Wilsonian flow equation. The latter must be solved simultaneously with Friedmann equations, which are modified by the $H$-dependence of the potential.

\vspace{5pt}
\paragraph*{Wilsonian flow equation.}
The coarse graining in space is characterised by a time-dependent scale $k$, which is eventually related to the Hubble rate.
Details of the derivation are given in the Supplemental Material, but we present here the main steps that led to our results. The inflaton is decomposed as 
\be
\Phi=\phi(t)+\varphi(t,\vec x)~,
\ee
where $\varphi$ is the quantum field, with three-dimensional Fourier transform defined with the comoving momentum $\vec p$
\be
\varphi(t,\vec p)=\int d^3x~\varphi(t,\vec x)e^{i\vec p\cdot\vec x}~.
\ee
Following Wetterich approach to Wilsonian renormalisation~\cite{Wetterich:1992yh,Wetterich:1989xg,Wetterich:2001kra}, but with a coarse graining in space,  we integrate over Fourier modes $\varphi(t,\vec p)$
weighted by a cutoff function $C_k(p)$ in the path integral, with $p=|\vec p|$. The scalar field bare action involved in the path integral is then
\bea\label{eq:action}
S_k[\Phi]&=&\int dt\sqrt{-g}\int d^3x\left(-\frac{g^{\mu\nu}}{2}\partial_\mu\Phi\partial_\nu\Phi-U_{bare}(\Phi)\right)\\
&&~~~~-\frac{1}{2}\int dt \sqrt{-g}\int\frac{d^3p}{(2\pi)^3}\varphi(t,\vec p)\varphi(t,-\vec p)~C_k(p)~,\nonumber
\eea
where the cutoff function plays the role of a mass of the order $k^2$ for IR modes (with $p^2<k^2$), 
and is suppressed for UV modes (with $p^2>k^2$). 
When defining the one-particle-irreducible (1PI) effective action $\Gamma_k$, UV modes are then integrated out without any constraint, 
while IR modes get frozen, leading to the {\em average effective action}~\cite{Wetterich:1989xg}. By construction, one imposes $\lim_{k\to0} C_k(p)=0$ which implies that, 
in the deep infrared limit $k\to0$, $\Gamma_k$ becomes the usual 1PI effective action.

Following Refs.~\cite{Wetterich:2001kra}, we compute this effective action using Eq.~\eqref{eq:action}, but for an arbitrary time-dependent cutoff function $k(t)$. We then perform a functional derivative of this effective action with respect to $k(t)$, and we obtain (see the Supplemental Material for a line-by-line derivation)
\bea\label{eq:dGammadk}
\frac{\delta\Gamma_k}{\delta k(t)}
&=&\frac{\sqrt{-g(t)}}{2}\int\frac{d^3p}{(2\pi)^3}\int\frac{d^3q}{(2\pi)^3}\delta(\vec p+\vec q\,)\nonumber\\
&&\int dt' \delta(t-t')
~{\cal O}(t,t',\vec p,\vec q\,)~\partial_k C_k(p)~,
\eea
where 
\bea\label{eq:O}
&&{\cal O}(t,t'\!,\vec p,\vec q\,)=\nonumber\\
&&\left[\!\sqrt{|g|}(C_k(p)-i\varepsilon)\delta(\vec p\!+\!\vec q\,)\delta(t\!-\!t')
-\frac{\delta^2\Gamma_k}{\delta\varphi_b(t,\vec p)\delta\varphi_b(t'\!,\vec q)}\right]^{\!-1}\!\!,\nonumber\\
\eea
and $\varphi_b$ stands for the vacuum expectation value of the quantum fluctuation $\varphi$.

We then consider the local potential approximation (LPA), where for every scale $k$ we assume that the effective action for the background $\phi$ has the form 
\be
\Gamma_k[\phi]=\int dt\sqrt{-g}\int d^3x\left(-\frac{g^{\mu\nu}}{2}\partial_\mu\phi\partial_\nu\phi-U_k(\phi)\right)~.
\ee
We use the Litim cutoff function \cite{Litim:2001up}, adapted to our situation, which reads
\be
C_k(p)=a^{-2}(k^2-p^2)\Theta(k^2-p^2)~.
\ee
In terms of the physical scale $\kappa\equiv k/a$, we find the Euclidean flow equation 
\be\label{dkappaU}
\partial_\kappa U_\kappa(\phi)=\frac{aT^{-1}}{6\pi^2}{\cal D}_E^{-1}\left(a^{-1}\kappa^4\right)~,
\ee
where 
\be
{\cal D}_E=-\frac{d^2}{dt^2}-3H\frac{d}{dt}+\kappa^2+\partial^2_\phi U_\kappa(\phi)~,
\ee
and $T^{-1}$ is a frequency cutoff, which is unavoidable since the cutoff function $C_k$ acts in three-dimensional Fourier space 
only\footnote{For usual four-dimensional Wilsonian flows, the cutoff function $C_k$ acts as a full regulator and no extra hard cutoff is needed.}.
This frequency cutoff  will be chosen as an initial Hubble scale $T^{-1}=H_0$. 

We then multiply by the operator ${\cal D}_E$ on both sides of Eq.~(\ref{dkappaU}), go back to real time and choose $\kappa=H$, to finally obtain the flow equation
\be\label{flowF}
\ddot F+H\dot F+(\partial^2_\phi U-H^2-\dot H)F=\frac{H_0H^4}{6\pi^2}~,
\ee
where $F\equiv\partial_H U$ and a dot denotes a total time derivative.

\vspace{5pt}
\paragraph*{Modified Friedmann equations.}
The action describing gravity and matter is 
\bea
\Sigma&=&\int\! d^4\!x\sqrt{-g}\left(\frac{M_p^2}{2}R-\frac{1}{2}\partial_\mu\phi\partial^\mu\phi-U_H(\phi)\right)\\
&=&\int\left\{\frac{3}{2}M_p^2\left[-\frac{2}{g_{00}}\left(\frac{\ddot a}{a}+\left(\frac{\dot a}{a}\right)^2\right)
+\frac{\dot g_{00}}{g_{00}^2}\frac{\dot a}{a}\right]\right.\nn
&&\left.~~~~~~-\frac{1}{2g_{00}}(\dot\phi)^2-U_H(\phi)\right\}~\sqrt{-g_{00}}~a^3~d^4x~,\nonumber
\eea
and the field equations are obtained from the variation with respect to the dynamical variables $g_{00}, a, \phi$.
When taking the variation of the action with respect to $g_{00}$ before setting it to $-1$, we should not forget the covariant definition (\ref{defH}) 
of the Hubble rate which appears in the potential. Denoting $F\equiv\partial_HU$, we have then
\begin{itemize}
\item $\delta \Sigma/\delta\phi=0$ for the scalar field evolution 
\be\label{dynphi}
\ddot\phi+3H\dot\phi+\partial_\phi U=0~,
\ee
\item $\delta \Sigma/\delta g_{00}=0$ for the Hamiltonian constraint
\be\label{constraint}
3M_p^2H^2=\frac{1}{2}(\dot\phi)^2+U - HF~,
\ee
\item $\delta \Sigma/\delta a=0$ for the scale factor dynamical equation  
\be\label{dyna}
M_p^2\left(2\frac{\ddot a}{a}+H^2\right)+\frac{1}{2}(\dot\phi)^2-U + HF + \frac{\dot F}{3}=0~.
\ee
\end{itemize}
The last two equations lead to the dynamical equation for $H$
\be\label{dynH}
2M_p^2~\dot H=-(\dot\phi)^2-\frac{\dot F}{3}~.
\ee
If one takes into account $\dot U=\dot H F+\dot\phi \partial_\phi U$,
one can check that the two Friedman equations (\ref{constraint}) and (\ref{dynH}) imply the scalar field equation (\ref{dynphi}), as expected, 
which is a consistency check for our approach. Trading the cosmic time variable for the number of $e$-folds $N\equiv \ln(a)$, with primes denoting now derivatives with respect to $N$, the master set of equations we need to solve is
\noindent\refstepcounter{equation}\label{eq:ODE}%
\begin{tcolorbox}[enhanced, colback=white, colframe=black, boxrule=0.5pt,
                  left=3pt, right=3pt, top=-4pt, bottom=2pt]
  \[
  \begin{aligned}
      F''&=\frac{H_0H^2}{6\pi^2}-\left(1+\frac{H'}{H}\right)F'
           -\left(\frac{\partial^2_\phi U}{H^2}-\frac{H'}{H}-1\right)F\,,\\
      \phi''&=-\left(3+\frac{H'}{H}\right)\phi'-\frac{\partial_\phi U}{H^2}\,,\\
      H'&=-\frac{H}{2M_p^2}\left((\phi')^2+\frac{F'}{3H}\right)\,.
  \end{aligned}
  \tag*{\((\theequation)\)} 
  \]
\end{tcolorbox}
Before we apply this formalism to describing cosmic inflation, let us now search for solutions of this system that lead to a de Sitter universe where all the $(\cdot)'$ derivatives in this system vanish.

\vspace{5pt}
\paragraph*{Cosmological Constant.}
From Eq.~(\ref{eq:ODE}), a flat potential $U=U_{\Lambda}$ and $H=H_0$ constant leads to
\be\label{eq:FlowdS}
H_0 F_{\Lambda}=-\frac{H_0^4}{6\pi^2}
\ee
Both Friedman equations are identical and lead to
\be
H_0F_{\Lambda}=U_{\Lambda}-3M_p^2H_0^2~,
\ee
such that 
\be\label{relationUHT_1}
U_{\Lambda}=3M_p^2H_0^2-\frac{H_0^4}{6\pi^2}~.
\ee
This case is trivial, but representative of what happens physically: for a given value of the cosmological constant $U_\Lambda$, the Hubble scale is given by Eq.~\eqref{relationUHT_1}, and one needs to adjust the flow of the cosmological constant with the Hubble scale ($F_\Lambda\not =0$) using Eq.~\eqref{eq:FlowdS} such that all the time derivatives in the master equation are zero and $H$ can remain constant. 

\vspace{5pt}
\paragraph*{de Sitter Solution.}
In the case where a scalar field would sit at the extremum of its potential ($\partial_\phi U=0$), demanding for a de Sitter universe imposes that
\be\label{eq:quasidS}
H_0F_{\rm dS}=-\frac{H_0^4}{6\pi^2(1-\partial_\phi^2U_{\rm dS}/H_0^2)}\,,
\ee
and 
\be\label{relationUHT}
U_{\rm dS}=3M_p^2H_0^2-\frac{H_0^4}{6\pi^2(1-\partial_\phi^2U_{\rm dS}/H_0^2)}~.
\ee
This last equation provides a relation between the potential $U_{\rm dS}$, its local curvature, and the initial Hubble scale $H_0$ in the de Sitter case.

\vspace{5pt}
\paragraph*{From de Sitter to Slow-Roll Inflation.}
Although this solution is a fixed point of the set of equations describing the dynamics, this equilibrium does not need to be stable. Indeed, in the case where the curvature of the potential is negative, quantum fluctuations might naturally push the field away from the de Sitter phase, leading the dynamics to deviate from this ideal solution. This is what we will consider as an initial condition for inflation in the next section. Indeed, we consider that the dynamics starts on a potential that is sufficiently flat that it can start in the quasi-de Sitter conditions of Eqs.~\eqref{eq:quasidS} and \eqref{relationUHT}. The only element that pushes the system away from its de Sitter state is the presence of a small slope $\partial_\phi U\not = 0$, which immediately enforces that $\phi$ rolls, leading to $H^\prime<0$.

In this {\em letter}, we will restrict ourselves to the case of the E-model class of $\alpha$-attractors~\cite{Kallosh:2013hoa}, although the following analysis could in principle be generalised to any kind of single-field slow-roll inflation. The potential we consider, is defined, at the Hubble scale $H_0$, as
\be\label{eq:inflation_pot}
U(\phi, H_0) = V(\phi)=V_0\left(1-e^{\sqrt{\frac{2}{3\alpha}}\frac{\phi}{M_p}}\right)^{2}\,.
\ee
This class of potentials is particularly attractive, as it encapsulates many different inflation scenarios, such as Starobinsky inflation ($\alpha=1$)~\cite{Starobinsky:1980te}, scenarios from
$\mathcal N = 4$ SUGRA~\cite{Cremmer:1978su4,Bergshoeff:1981ecsg,deRoo:1985matter} ($\alpha=1/3$), or so-called GL models~\cite{Ferrara:2012,Goncharov:1984JETP,Goncharov:1984PLB,Linde:2015jcap} ($\alpha=1/9$).

\vspace{5pt}
\paragraph*{{\bf Numerical analysis ---}}
The set of Eqs.~\eqref{eq:ODE} is, in principle, infinite-dimensional. Indeed, calculating partial derivatives of the potential $U_H(\phi)$ with respect to $\phi$ would require solving this system of non-linear differential equations for multiple trajectories $\phi(N)$ and computing on a lattice partial derivatives at every point. Instead, one can use a decomposition of the potential on a basis that helps tracking the evolution of the potential along the trajectory of interest. Before we present our results, we detail here our full numerical procedure.

\vspace{5pt}
\paragraph*{Exponential Basis Decomposition.}
One possibility of such decomposition, given the form of the potential \eqref{eq:inflation_pot}, is to use an exponential basis
\be
U_H(\phi)=\sum_{n=0}^\infty U_n(H)~e^{-n\phi/M}~,
\ee
where we defined
 $M=\sqrt{3\alpha/2}~M_p$. The corresponding variable $F$ is then written
\be
F=\sum_{n=0}^\infty F_n~e^{-n\phi/M}~,~\mbox{with}~~F_n=\frac{dU_n}{dH}~.
\ee
The exponential expansions above are plugged into the flow equation (first line of Eq.~(\ref{eq:ODE})), and the identification of each power of the exponential leads to
\bea\label{dynFn}
F_n'' &=& \delta_{n,0} \frac{H_0H^2}{6\pi^2}-F_n'\left[-2n\frac{\phi'}{M}+\left(1+\frac{H'}{H}\right)\right]\nonumber\\
&&+ F_n \left[n\frac{\phi''}{M}-\left(n\frac{\phi'}{M}\right)^2+\left(1+\frac{H'}{ H}\right)\left(1+n\frac{\phi'}{M}\right)\right]\nonumber\\
&&- \frac{1}{H^2M^2}\left(\sum_{l=0}^n l^2U_l F_{n-l}\right)~,\qquad n\geqslant 0.
\eea
After truncating this ladder of equations to a sufficiently high order to ensure numerical stability, it is to be solved simultaneously with the Friedmann equations for $\phi(N)$ and $H(N)$ in Eq.~\eqref{eq:ODE}.

\vspace{5pt}
\paragraph*{Initial Conditions.}
For the potential of Eq.~\eqref{eq:inflation_pot} and a given set of potential parameters ($V_0,\alpha$), the initial values of the $U_n$ is determined at all orders:
\be
U_0=V_0\,,\quad U_1=-2V_0\,,\quad U_2=V_0\,,\quad U_{n>2}=0\,.
\ee
After fixing the initial value of the field, $\phi_0$, we then  proceed in two steps: First, we use Eq.~\eqref{relationUHT} to obtain the value of the Hubble scale at initial time, $H_0$, then we perform the same exponential decomposition described above of Eq.~\eqref{eq:quasidS}, giving
\be
F_n(0) = \left\{\begin{matrix}
    \displaystyle -\frac{H_0^3}{6\pi^2} & \qquad \text{if $n=0$}\\ 
    ~\\
    \displaystyle \sum_{l=0}^n \frac{l^2U_l(0) F_{n-l}(0) }{H_0^2M^2}& \qquad \text{if $n\geqslant 1.$}
\end{matrix}\right.
\ee
Together, these equations guarantee that the Universe starts in an instant de Sitter state defined by $\phi'=H'=F'=0$, and only deviates from its de Sitter state because of the field quickly reaching its slow-roll attractor.

\begin{figure}
    \centering
    \includegraphics[width=\linewidth]{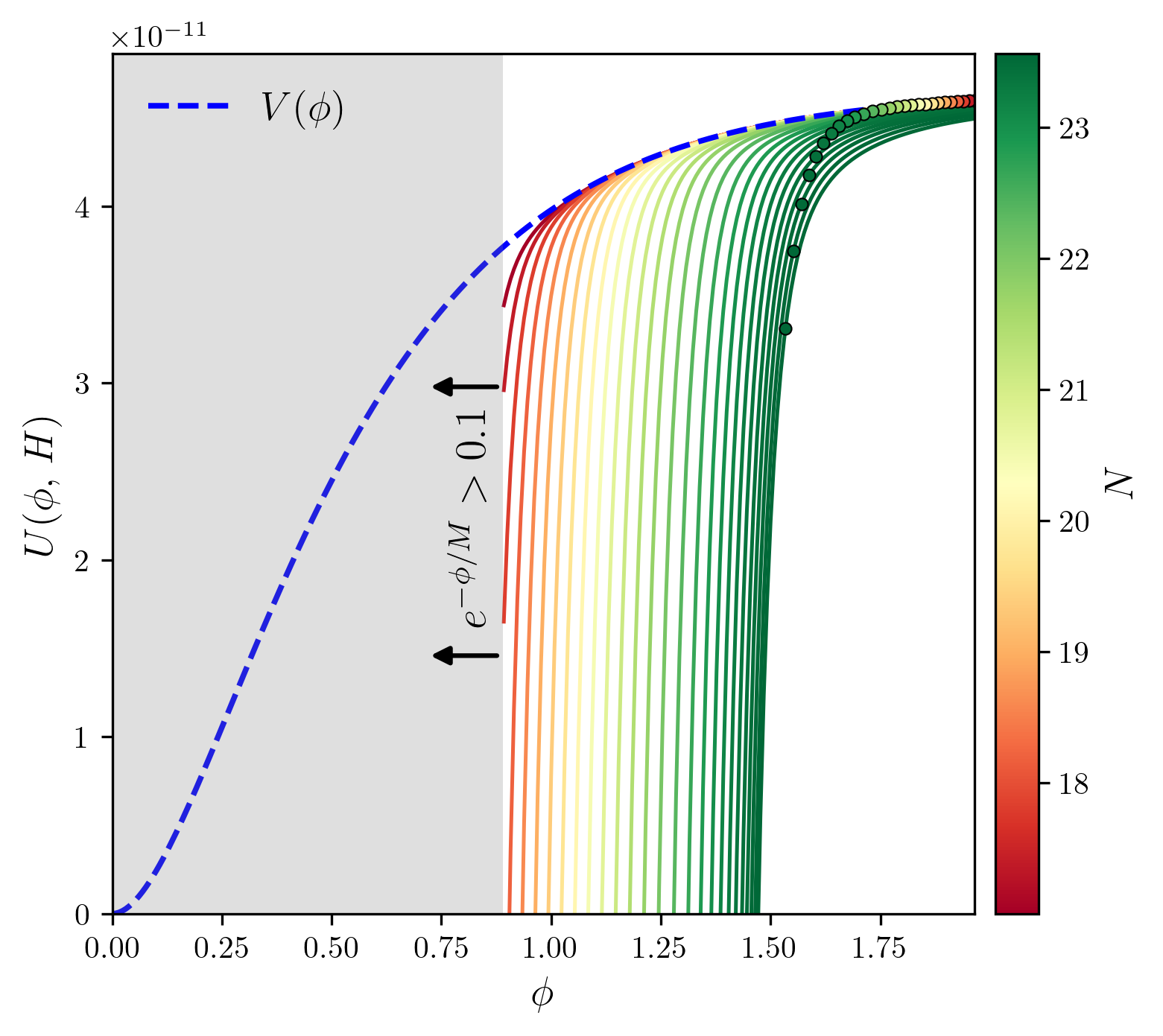}
    \caption{\label{fig:example} \footnotesize Full ERG evolution of the potential $U(\phi,H(N))$ (coloured lines), for $\alpha=0.1$, $\phi_0=2.3M_p$, and for various values of the number of $e$-folds since horizon crossing (assumed to be at $\phi(0)=\phi_0$). Coloured dots depict the corresponding values of the solution $\phi(N)$ when these potentials are evaluated.}
\end{figure}

\vspace{5pt}
\paragraph*{Effect on the Inflationary Dynamics.}
A representative example of the inflationary dynamics and its ERG flow evolution is represented in FIG.~\ref{fig:example}, in the case where $\alpha=0.1$ and $\phi_0=2.3M_p$. In this figure, both the Hubble scale and field value evolve with time (or the number of $e$-folds) while the scalar potential is driven by flow equation, accordingly to the set of equations~\eqref{eq:ODE}. The solution $\phi(N)$ is depicted with coloured dots, while coloured curves indicate the overall shape of the potential when evaluated at the corresponding values of $H(N)$ for arbitrary $\phi$. In that plot, the grey-shaded area stands for the region where $e^{-\phi/M}>0.1$ and we no longer trust the exponential decomposition described, used to solve equations numerically. As one can see from the figure, while the field rolls down, the potentials drop at larger and larger field values over time, thus destabilising the potential and prematurely accelerating the field, enforcing slow-roll exit earlier than in the absence of running. Remarkably, this has the important effect of pushing the field value at which the CMB modes exit the horizon further away on the plateau of the potential, modifying drastically the inflation observables, as we will see in the next subsection.
\begin{figure*}    \includegraphics[width=\linewidth]{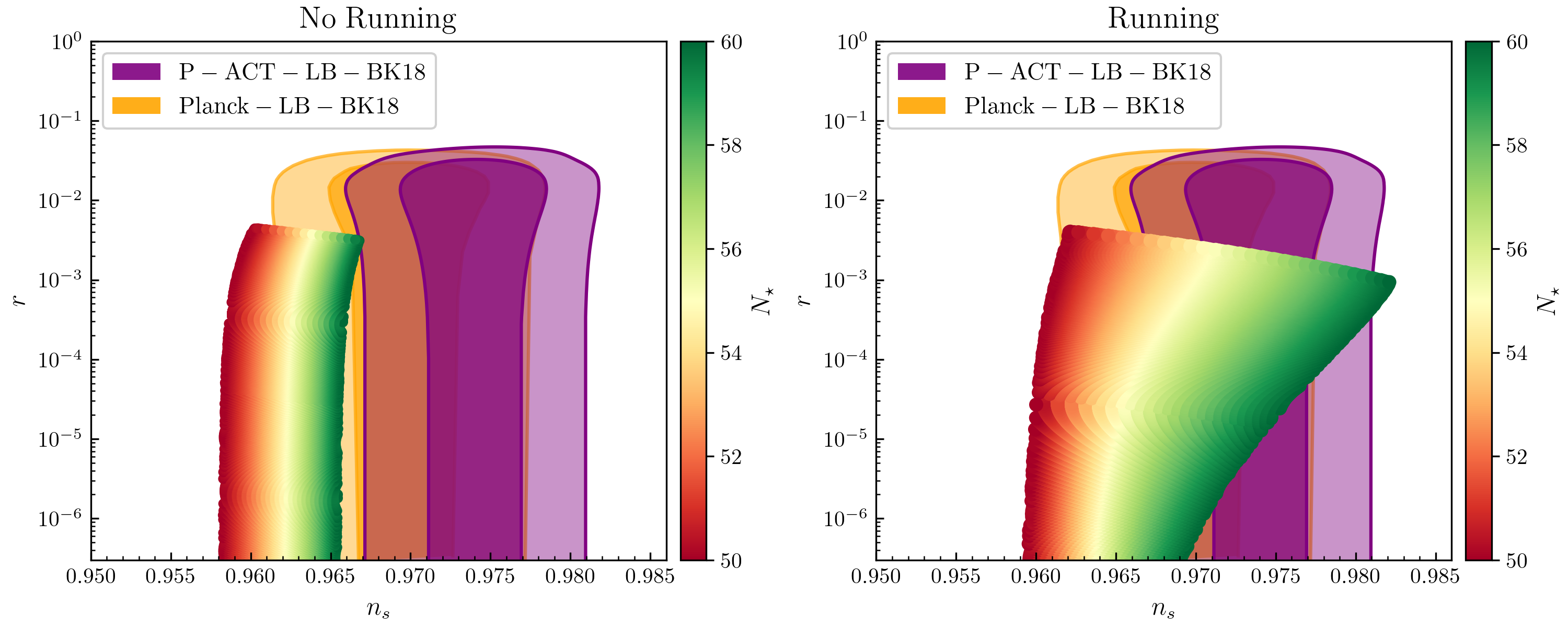}
    \caption{\label{fig:Emodel}\footnotesize Tensor-to-scalar ratio $r$ as a function of the spectral index $n_s$ calculated with (right panel) or  without (left panel) including the ERG flow of the potential.}
\end{figure*}

\vspace{5pt}
\paragraph*{Inflation Observables.}
To explore the effect of the inflationary ERG flow on CMB observables, we scan over the parameter space for various values of $\alpha\in[10^{-5},1]$ and $\phi_0\in[0.05M_p,7M_p]$. For each parameter choice, we calculate the slow-roll parameters
\be
\epsilon_V\equiv \frac{M_p^2}{2}\left(\frac{\partial_\phi U}{U}\right)^2\,,\quad \eta_V\equiv M_p^2\frac{\partial_\phi^2 U}{U}\,,
\ee
and normalise the potential at $\phi=\phi_0$ such that the primordial power spectrum of scalar perturbations has the correct amplitude at the {\it Planck} pivot scale~\cite{Planck:2018vyg}
\be
A_s \equiv \frac{U(\phi_0,H_0)}{24\pi^2\epsilon_V(\phi_0)}\approx 2.1\times 10^{-9}\,.
\ee
Then we track the dynamics of the inflaton, Hubble scale, flow functions $F_n$, and potential functions $U_n$ dynamically until the system exits slow-roll when $\varepsilon_H\equiv-{H'}/{H} = 1$.
We then calculate $N_\star$, the number of $e$-folds between horizon exit and the end of inflation, and select parameters for which $50<N_\star<60$. The tensor-to-scalar ratio and spectral index of the primordial scalar perturbation power spectrum are then evaluated using their slow-roll  expression at horizon exit, 
$r\approx 16\epsilon_V(\phi_0)$ and \mbox{$n_s\approx 1-6\epsilon_V(\phi_0)+2\eta_V(\phi_0)$}.

\vspace{5pt}
\paragraph*{Results.}
Our results are presented in FIG.~\ref{fig:Emodel}, in which the left panel stands for the usual case without running, and the right panel included the full ERG flow in the dynamics. In the latter case, we pushed the expansion to $n=15$ and verified that our results are insensitive to that choice. Results are compared to the most recent constraints from {\it Planck}/BICEP/Keck~\cite{BICEP:2021xfz} and ACT~\cite{ACT:2025tim}. As expected from the previous qualitative discussion, the effect of the running is to provoke the ending of inflation much earlier compared to the case of a rigid potential. This effectively pushes the value of the field at horizon crossing to larger values where the potential is flatter, effectively increasing the value of the spectral index.

\vspace{5pt}
\paragraph*{{\bf Summary and Discussion ---}}
In this paper, we have explored the evolution of the inflationary dynamics in the presence of an evolving effective potential. Due to the  quantum modes exiting the horizon during inflation, and hence stopping contributing to the effective potential felt by the classical inflaton background field, this evolution ultimately destabilises the slow-roll regime, enforcing an early ending of inflation, and effectively pushing the inflaton to be further up its potential when the CMB modes exit the horizon. Using the example of $\alpha$-attractor E-models, we saw that the effect of this ERG running is to reconcile the model with recent CMB data from ACT that were initially thought to exclude $\alpha$-attractor models with 95\% confidence level.

While this work proposes a new framework for the study of inflation, it also opens several new avenues that ought to be explored in the future. First of all, we ignored in this work the backreaction of metric fluctuations on the inflaton dynamics. Whereas this is a good approximation deep inside the slow-roll regime~\cite{George:2013iia}, incorporating such fluctuations may lead to modification of the dynamics when ERG effects become important. In fact, it is visible in our analysis that the destabilisation of the inflaton potential cannot continue forever, as it seems to lead to a divergence that is not physical, but may instead be assimilated to a {\em spinodal} instability well-known in the context of Euclidean ERG flow equations
. 
For these reasons, it is apparent that this framework cannot be used to properly describe the behaviour of the field after inflation ends, and it is possible that including metric fluctuations in the calculation---or using alternative regularisation methods that do not feature explicit spinodal instabilities~\cite{Abel:2023ieo} or proposes an alternative Wilsonian renormalisation scheme relevant to cosmology~\cite{Marian:2019nhu}---may help pushing our understanding of renormalisation beyond the sole study of slow-roll inflation. Finally, our treatment does not account for the stochastic effect exerted by quantum modes that exit the horizon on the classical background field characteristic of open-EFT systems. Upgrading this framework using the closed-time-path formalism~\cite{Schwinger:1960qe,Keldysh:1964ud,Calzetta:1986ey,Salcedo:2024smn} is thus something we are planning to explore in  following studies. 

By dynamically incorporating the effect of quantum corrections into the description of the Universe's evolution, our work sets the path for a new approach to scalar theories of the early universe. This may lead the community to reconsider the impact of cosmic inflation theories on cosmological observables, as well as broader classes of theories, such as quintessence scenarios, early dark energy models, or even axion-like particle phenomenology.

\section*{Acknowledgements}
The authors would like to thank Wenyuan Ai, Bj\"orn Garbrecht, Thomas Colas, and Enrico Pajer for stimulating discussions.  The work of JA and LH was supported by the STFC under UKRI grant ST/X000753/1. The work of SP was funded by the Deutsche Forschungsgemeinschaft (DFG, German Research Foundation) under Germany’s Excellence Strategy – EXC 2094 – 390783311. LH would like to thank the Institute for Particle Physics Phenomenology (IPPP) at Durham University for providing us with access to its high-performance computing resources, which facilitated the realisation of this work.

\vfill

\bibliographystyle{apsrev4-1}
\bibliography{main}

\clearpage

\onecolumngrid

\begin{center}
\textbf{\Large Supplemental Material
}
\end{center}
\noindent

\renewcommand{\thesection}{S\arabic{section}}
\renewcommand{\thefigure}{S\arabic{figure}}
\renewcommand{\theequation}{S\arabic{equation}}

\setcounter{equation}{0}
\setcounter{page}{1}
\setcounter{figure}{0}


\section{Flow equation: Detailed Derivation}\label{App:flow}

The total scalar field is $\Phi=\phi(t)+\varphi(t,\vec x)$, and the partition function with 3-dimensional coarse-graining is defined 
for the quantum field $\varphi$ on the background $\phi$:
\bea \label{eq:eq00}
Z_k[j]&\equiv& e^{iW_k[j]}\\
&=&\int{\cal D}[\varphi]\exp\left(iS[\Phi]+i\int dt \sqrt{-g}\int d^3x ~j\varphi\right.\nn
&&\left.-\frac{i}{2}\int dt \sqrt{-g}\int\frac{d^3p}{(2\pi)^3}\varphi(t,\vec p)\varphi(t,-\vec p)(C_k(p)-i\varepsilon)\right) ~,\nonumber
\eea
where $\varepsilon>0$. The "$i\varepsilon$" prescription ensures that the path integral converges, and will avoid a pole in the integration over space momentum, as explained further down. The background fluctuation field with momentum $\vec p$ is
\be
\varphi_b(t,\vec p)=\left<\varphi(t,\vec p)\right>
\equiv\frac{1}{\sqrt{-g}}\frac{\delta W_k}{\delta j(t,-\vec p)}=-\frac{i}{\sqrt{-g}Z_k}\frac{\delta Z_k}{\delta j(t,-\vec p)}~,
\ee
and the second functional derivative of $W_k$ is 
\be
\frac{\delta^2 W_k}{\delta j(t,-\vec p)\delta j(t',-\vec q)}
=i\sqrt{g(t)g(t')}\left[\left<\varphi(t,\vec p)\varphi(t',\vec q)\right>-\varphi_b(t,\vec p)\varphi_b(t',\vec q)\right]~.
\ee
The one-particle-irreducible effective action is defined as the Legendre transform of $W[j]$, where the terms involving $\varphi_b$ only cancel out
\bea
\Gamma_k[\phi+\varphi_b]&=&W_k[j]-\int dt' \sqrt{-g}\int d^3x ~j\varphi_b \\
&&+\frac{1}{2}\int dt' \sqrt{-g}\int\frac{d^3p}{(2\pi)^3} \varphi_b(t',\vec p)\varphi_b(t',-\vec p)(C_k(p)-i\varepsilon)~,\nonumber
\eea
and $j$ should be understood as a functional of $\varphi_b$. $\Gamma_k$ has the following functional derivatives
\bea
\frac{1}{\sqrt{-g}}\frac{\delta\Gamma_k}{\delta\varphi_b(t,p)}&=&-j(t,-\vec p)+\varphi_b(t,-\vec p)(C_k(p)-i\varepsilon)\\
\frac{\delta^2\Gamma_k}{\delta\varphi_b(t,\vec p)\delta\varphi_b(t',\vec q)}&=&\sqrt{-g(t)}(C_k(p)-i\varepsilon)\delta(\vec p+\vec q)\delta(t-t')\nn
&&-\sqrt{g(t)g(t')}\left(\frac{\delta^2W_k}{\delta j(t,-\vec p)\delta j(t',-\vec q)}\right)^{-1}~.\nonumber
\eea
The evolution equation with $k(t)$ is then
\bea\label{dGammadk}
&&\frac{\delta\Gamma_k}{\delta k(t)}\\
&=&\frac{\delta W}{\delta k(t)}
+\frac{\sqrt{-g(t)}}{2}\int\frac{d^3p}{(2\pi)^3} \varphi_b(t,\vec p)\varphi_b(t,-\vec p)\partial_k C_k(p)\nn
&=&\frac{\sqrt{-g(t)}}{2}\int\frac{d^3p}{(2\pi)^3} 
\Big(\varphi_b(t,\vec p)\varphi_b(t,-\vec p)-\left<\varphi(t,\vec p)\varphi(t,-\vec p)\right>\Big)\partial_k C_k(p)\nn
&=&i\frac{\sqrt{-g(t)}}{2}\int\frac{d^3p}{(2\pi)^3}\int\frac{d^3q}{(2\pi)^3}\delta(\vec p+\vec q)\int dt' \delta(t-t')
~{\cal O}(t,t',\vec p,\vec q)~\partial_k C_k(p)~,\nonumber
\eea
where 
\bea\label{O}
&&{\cal O}(t,t',\vec p,\vec q)\\
&=&\frac{1}{\sqrt{g(t)g(t')}}\frac{\delta^2W_k}{\delta j(t,-\vec p)\delta j(t',-\vec q)}\nn 
&=&\left(\sqrt{-g(t)}(C_k(p)-i\varepsilon)\delta(\vec p+\vec q)\delta(t-t')
-\frac{\delta^2\Gamma_k}{\delta\varphi_b(t,\vec p)\delta\varphi_b(t',\vec q)}\right)^{-1}~.\nonumber
\eea

Next we assume the local potential approximation (LPA) where, for every scale $k$, we have
\bea
&&\Gamma_k[\phi+\varphi_b]\\
&=&\int dt\sqrt{-g}\int d^3x\left(-\frac{g^{\mu\nu}}{2}\partial_\mu(\phi+\varphi_b)\partial_\nu(\phi+\varphi_b)-U_k(\phi+\varphi_b)\right)\nn
&=&\int dt\sqrt{-g}\int\frac{d^3p}{(2\pi)^3}
\left(\frac{1}{2}\big(\dot\phi(t)+\dot\varphi_b(t,\vec p)\big)\big(\dot\phi(t)+\dot\varphi_b(t,-\vec p)\big)-\frac{p^2}{2a^2}\varphi_b(t,\vec p)\varphi_b(t,-\vec p)\right)\nn
&&~~~~~~~-\int dt\sqrt{-g}\int d^3x~ U_k(\phi+\varphi_b)~,\nonumber
\eea
such that, for a vanishing source $j=0=\varphi_b$
\be\label{dGammadkbis}
\frac{1}{\sqrt{-g}}\frac{\delta\Gamma_k[\phi]}{\delta k(t)}=-V~\partial_kU_k(\phi)~,
\ee
where $V$ is the space volume. Also, in the LPA, the operator (\ref{O}) is 
\bea\label{Otpq}
&&{\cal O}(t,t',\vec p,\vec q)=\delta(t-t')\delta(\vec p+\vec q)\\
&&~~~~\times\Big(\partial_t(\sqrt{-g}\partial_t)+\sqrt{-g}\left[p^2a^{-2}+U_k''(\phi)+C_k(p^2)-i\varepsilon\right]\Big)^{-1}~.\nonumber
\eea 
For the previous step we used the fact that the distributions $\delta(t-t')$ and $\delta(\vec p+\vec q)$ are identical to their inverse.
The derivatives $\partial_t(\sqrt{-g}~\partial_t)$ acting on the time-dependent quantities, when taking the inverse, 
introduces poles in the complex $p$-plane. To avoid this, we perform the Wick rotation $t\to it$ 
and, in the limit $\varepsilon\to0$, the resulting Euclidean version of the operator (\ref{Otpq}) acting on the cutoff function is
\be
{\cal O}_E(t,t',\vec p,\vec q)\partial_k C_k(p^2)=\delta(t-t')\delta(\vec p+\vec q){\cal D}_E^{-1}\left(a^{-3}\partial_k C_k(p^2)\right),
\ee
where the Euclidean differential operator is
\be
{\cal D}_E=-a^{-3}\frac{d}{dt}\left(a^3\frac{d}{dt}\right)+p^2a^{-2}+\partial^2_\phi U_k(\phi)+C_k(p^2)~.
\ee
After the Wick rotation, Eq.~(\ref{dGammadkbis}) should involve the Euclidean effective action
\be
\frac{1}{\sqrt{-g}}\frac{\delta\Gamma_k[\phi]}{\delta k(t)}\to 
\frac{i}{\sqrt{g}}\frac{\delta\Gamma_k^E[\phi]}{\delta k(t)}=iV~\partial_k U_k(\phi)~,
\ee
such that the evolution equation for the potential is finally
\be\label{dkU}
\partial_k U_k(\phi)=\frac{T^{-1}}{2}\int\frac{d^3p}{(2\pi)^3}{\cal D}_E^{-1}\Big(a^{-3}\partial_kC_k(p^2)\Big)~,
\ee
where $T^{-1}\equiv \lim_{t'\to t}\delta(t-t')$ is a cut-off frequency and we used the identity $\lim_{\vec p+\vec q\to0}\delta(\vec p+\vec q)=V$.\\

For the next steps we choose the following adaptation from the Litim cutoff
\be
C_k(p)=a^{-2}(k^2-p^2)\Theta(k^2-p^2)~,
\ee
and we introduce the physical scale $\kappa\equiv k/a$, in which case the evolution equation becomes 
\be\label{flowE}
\partial_\kappa U_\kappa(\phi)=\frac{aT^{-1}}{6\pi^2}{\cal D}_E^{-1}\left(a^{-1}\kappa^4\right)~~~~~~(\kappa=k/a)~,
\ee
where 
\be
{\cal D}_E=-\frac{d^2}{dt^2}-3H\frac{d}{dt}+\kappa^2+\partial^2_\phi U_\kappa(\phi)~.
\ee
Eq.(\ref{flowE}) can also be written
\be\label{step}
a{\cal D}_E\big(a^{-1}\partial_\kappa U\big)=\frac{T^{-1}\kappa^4}{6\pi^2}~,
\ee
which, after going back to real time, becomes
\be\label{flow}
\frac{d^2(\partial_\kappa U)}{dt^2}+H\frac{d(\partial_\kappa U)}{dt}+(\kappa^2+\partial^2_\phi U-2H^2-\dot H)(\partial_\kappa U)=\frac{T^{-1}H^4}{6\pi^2}~.
\ee
If we take $\kappa=H$ we finally obtain
\be\label{finalflow}
\ddot F+H\dot F+(\partial^2_\phi U-H^2-\dot H)F=\frac{T^{-1}H^4}{6\pi^2}~,
\ee
with $F\equiv\partial_H U$.

\end{document}